\documentclass[aps,prl,amsmath,amssymb,reprint,superscriptaddress,longbibliography]{revtex4-1}

\usepackage{graphicx}  
\usepackage{bm}        
\usepackage{amssymb}   
\usepackage{lettrine}
\usepackage{soul}
\usepackage{color}
\usepackage{braket}
\usepackage{tabularx}
\usepackage{subcaption}
\usepackage{mathtools}
\usepackage{verbatim}
\usepackage{amsmath}
\usepackage{caption}
\usepackage{subcaption}

\begin{document}

\title{Low temperature properties of whispering gallery modes in isotopically pure silicon-28}
	
\author{J. Bourhill}
\email{jeremy.bourhill@uwa.edu.au}
\author{M. Goryachev}

\affiliation{ARC Centre of Excellence for Engineered Quantum Systems, University of Western Australia, 35 Stirling Highway, Crawley WA 6009, Australia}

\author{D.L. Creedon}
\affiliation{School of Physics, University of Melbourne, Victoria 3010, Australia}

\author{B.C. Johnson}
\author{D.N. Jamieson}
\affiliation{Centre for Quantum Computation and Communication Technology,
School of Physics, University of Melbourne, Parkville, VIC 3010, Australia}

\author{M.E. Tobar}
\affiliation{ARC Centre of Excellence for Engineered Quantum Systems, University of Western Australia, 35 Stirling Highway, Crawley WA 6009, Australia}
\date{\today}


\begin{abstract}
\noindent {Whispering Gallery (WG) mode resonators have been machined from a boule of single-crystal isotopically pure silicon-28. Before machining, the as-grown rod was measured in a cavity, with the best Bragg confined modes exhibiting microwave $Q$-factors on the order of a million for frequencies between 10 and 15 GHz. After machining the rod into smaller cylindrical WG mode resonators, the frequencies of the fundamental mode families were used to determine the relative permittivity of the material to be $11.488\pm0.024$ near 4 K, with the precision limited only by the dimensional accuracy of the resonator. However, the machining degraded the $Q$-factors to below $4\times10^4$. Raman spectroscopy was used to optimize post-machining surface treatments to restore high $Q$-factors. This is an enabling step for the use of such resonators for hybrid quantum systems and frequency conversion applications, as silicon-28 also has very low phonon losses, can host very narrow linewidth spin ensembles and is a material commonly used in optical applications.}
\end{abstract}

\maketitle

\section{Introduction}
In this work we describe investigations of the microwave properties of isotopically purified silicon-28 ($^{28}$Si). The dielectric properties of the material are close to that of naturally occurring silicon, which is the primary material used in the manufacturing of semiconductor devices. Standard silicon is a mixture of silicon atoms consisting of 28, 29 and 30 nucleons with approximately $92.2\%$, $4.7\%$ and $3.1\% $ abundance in nature, respectively. Of these isotopes, only $^{28}$Si and $^{30}$Si have zero nuclear spin. This feature means that pure crystals of these isotopes act as a so-called `virtual vacuum' with respect to a bath of magnetic Two Level Systems (TLS), leaving the medium inert \cite{Wolfowicz2013}, which is not true for standard silicon. Thus, the absence of nuclear spin in the lattice circumvents one of the most important channels of decoherence, leaving only phonon dissipation which can be greatly reduced through cooling.

As a result of the lack of nuclear spin, enriched $^{28}$Si has been demonstrated to be an ideal host material for solid-state qubits. Electron or nuclear spins of low-concentration impurities in bulk $^{28}$Si can form extremely well isolated quantum systems. These systems have been shown to exhibit coherence times ($T_2$) of up to seconds for the electron spin \cite{spintime}, and minutes for the nuclear spin \cite{nspintime}, which are comparable to those of qubits based on trapped ions \cite{trapped1,trapped2}. In addition, isotopically purified silicon has the potential to allow the development of a solid-state clock formed from an ensemble of impurity donors\cite{Saeedi2015}, a scheme which has also been proposed for other solid-state mediums\cite{Jin:2015fr,Hodges2013,Goryachev2018}. The measurement of clock transitions may make use of a microwave resonant cavity. If the silicon crystal itself acted as the microwave resonant cavity, utilising a particularly low-loss resonant family of modes - whispering gallery (WG) modes, the frequency stability and interaction strengths (and hence coherence times) could be vastly improved in these clock transitions. Furthermore, by manufacturing these resonators from enriched silicon, dephasing as a result of nuclear spin interaction could be removed.

Previously, a very low-loss paramagnetic spin ensemble was detected in enriched silicon with a narrow linewidth of less than 7 kHz for a 10 parts per trillion concentration of impurity ions\cite{nikita}. This was only possible by virtue of the low dielectric photonic losses combined with the long lifetime of the spin transition (low magnetic loss), which enhanced the AC magnetic susceptibility. These results indicate that single crystal $^{28}$Si has immense potential for future cavity QED experiments and presents as an excellent candidate for a solid-state clock if impurities that exhibit clock transitions could be purposefully doped within the crystal.

\begin{figure}[b]
\includegraphics[width=\columnwidth]{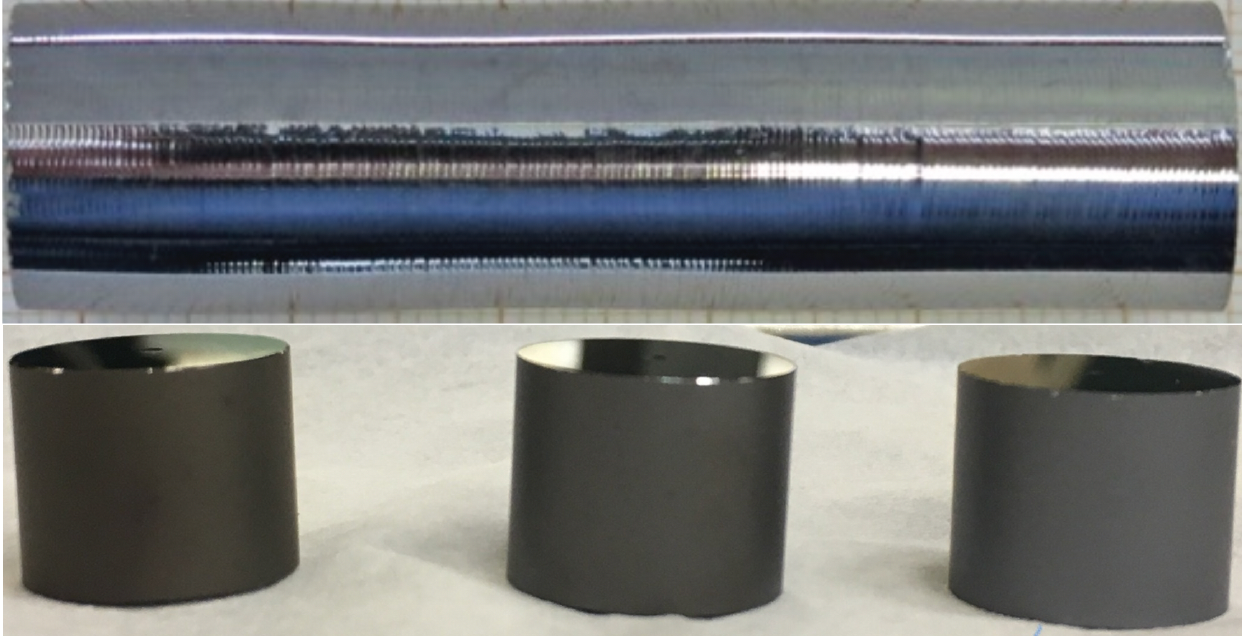}
\caption{(Top) Photograph of the $^{28}$Si rod of $55$ mm in length and approximately $15$ mm diameter before machining. (Below) Three of the four WG mode $^{28}$Si cylindrical  resonators cut and machined from the rod.}
\label{isopure}
\end{figure}

In this work we present an analysis of WG mode measurements in an enriched silicon resonator. We characterise the modes and their losses, and discuss the manufacturing and treatment techniques that have been used to improve them.

\begin{figure}[t]
\includegraphics[width=\columnwidth]{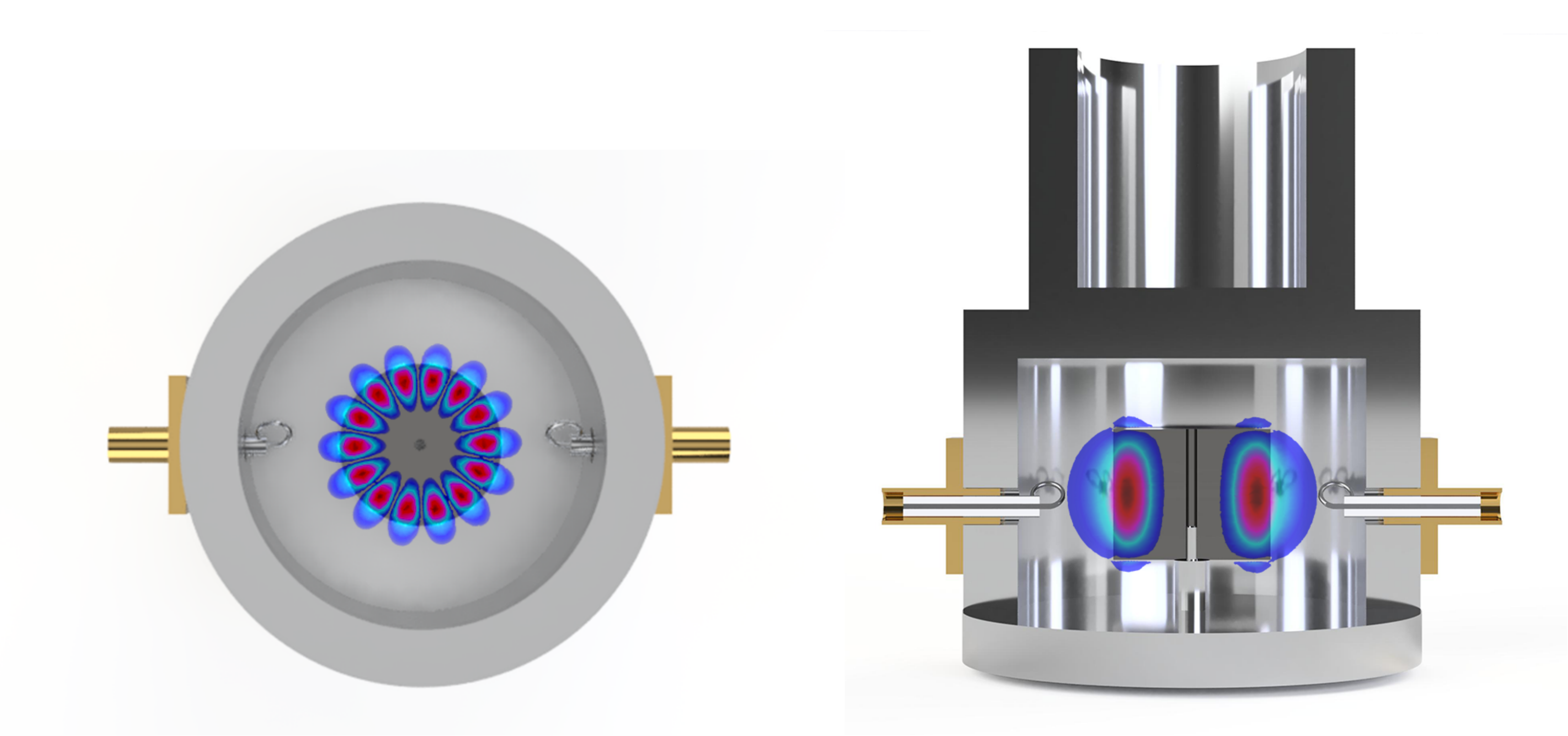}
\caption{Graphical representation of the cylindrical $^{28}$Si WG mode resonator loaded cavity. The $E_z$ field component of the $19.535$ GHz WGH$_{7,1,1}$ mode, calculated by Finite Element Analysis using COMSOL, is shown overlaid on the resonator. (Left) Cross-section through in the $r-\theta$ plane at $z = 0$.  (Right) Cross-section through the $r-z$ plane. The coaxial coupling probes used for excitation and measurement are shown, which couple to the evanescent field of the resonator.}
\label{comsol}
\end{figure}

\section{Preparation of isotopically pure whispering gallery mode resonators}
A crystal boule approximately $15$ mm diameter and $55$ mm length was grown using the float zone melting method in an argon atmosphere\cite{Itoh2003} at the G.G. Devyatykh Institute of Chemistry of High-Purity Substances. The specified concentration of $^{28}$Si was $99.993\%$, with concentration of oxygen and carbon estimated to be less than $1.1\times10^{16}$ cm$^{-3}$ (on the order of 1 ppm) and boron and phosphorous to be less than $1.1\times10^{13}$ cm$^{-3}$ (on the order of 1 ppb). The boule as-drawn from the melt is cylindrical, with a radius that can clearly be seen to vary by visual inspection. Thus, initial predictions of mode frequencies in the boule were imprecise, making it difficult to properly characterize any photon-spin interactions. To solve this problem, the boule was machined to make four smaller, more dimensionally precise cylindrical resonators. The crystal was turned to have a uniform cylindrical diameter, which was measured to be $14.455\pm0.010$ mm. This rod was then diced into four equal cylinders of 14.90 mm height, which were optically polished on the end faces and drilled through the axial centre of the crystal to a diameter of 1 mm. A photo of the as-grown rod along with three of the diced cylindrical resonators is shown in Fig. \ref{isopure}. Two cavities to house the WG mode resonators for characterization were then machined from oxygen-free copper and aluminium, each with an integrated support post.  A schematic of the resonator loaded cavity is shown in Fig. \ref{comsol}.

\section{Characterization of isotopically pure whispering gallery mode resonators}

The cylindrical WG mode resonators support quasi-Transverse Electric (WGE) and quasi-Transverse Magnetic (WGH) modes.  To allow accurate prediction of these mode frequencies, precise measurements of all dimensions of the cavity and resonator were first made at room temperature. We then calculated the dimensional change at cryogenic temperatures using published values for the coefficient of thermal expansion for all materials. The resulting dimensions were used to define a model in finite element analysis software (COMSOL Multiphysics), in which we implemented an ultra-fine mesh that reduced frequency errors to below that of the dimensional errors of the WG mode cavity. An example of the computed $\vec{E}_Z$ field density for the WGH$_{7,1,1}$ mode at $19.535$ GHz is shown in Fig. \ref{comsol}, overlayed on top of a 3D rendering of the crystal and housing. Initial simulations were performed using the low temperature permittivity value for standard float-zone silicon at microwave frequencies published previously\cite{jerzy}. That value, $\epsilon_r = 11.450\pm0.012$, was calculated from the frequencies of higher order TE$_{0,x}$ modes.

For measurement, the resonator loaded cavity was placed under vacuum and cooled to as low as 20 mK with a dilution refrigerator. The experimental setup was similar to that used previously for investigations of other materials at low temperature\cite{farr,giantg,singlephoton}. In this setup, the incident microwave signal was attenuated to the level of a few photons by a chain of room temperature and cold attenuators, and the output signal was amplified by a series of cold high-electron-mobility transistor (HEMT) and room temperature amplifiers. These signals were coupled to the resonator using diagonal loop probes to permit coupling to both transverse and axial $\vec{H}$ fields, and hence to both WGH and WGE families.

\begin{figure}[t]
\includegraphics[width=\columnwidth]{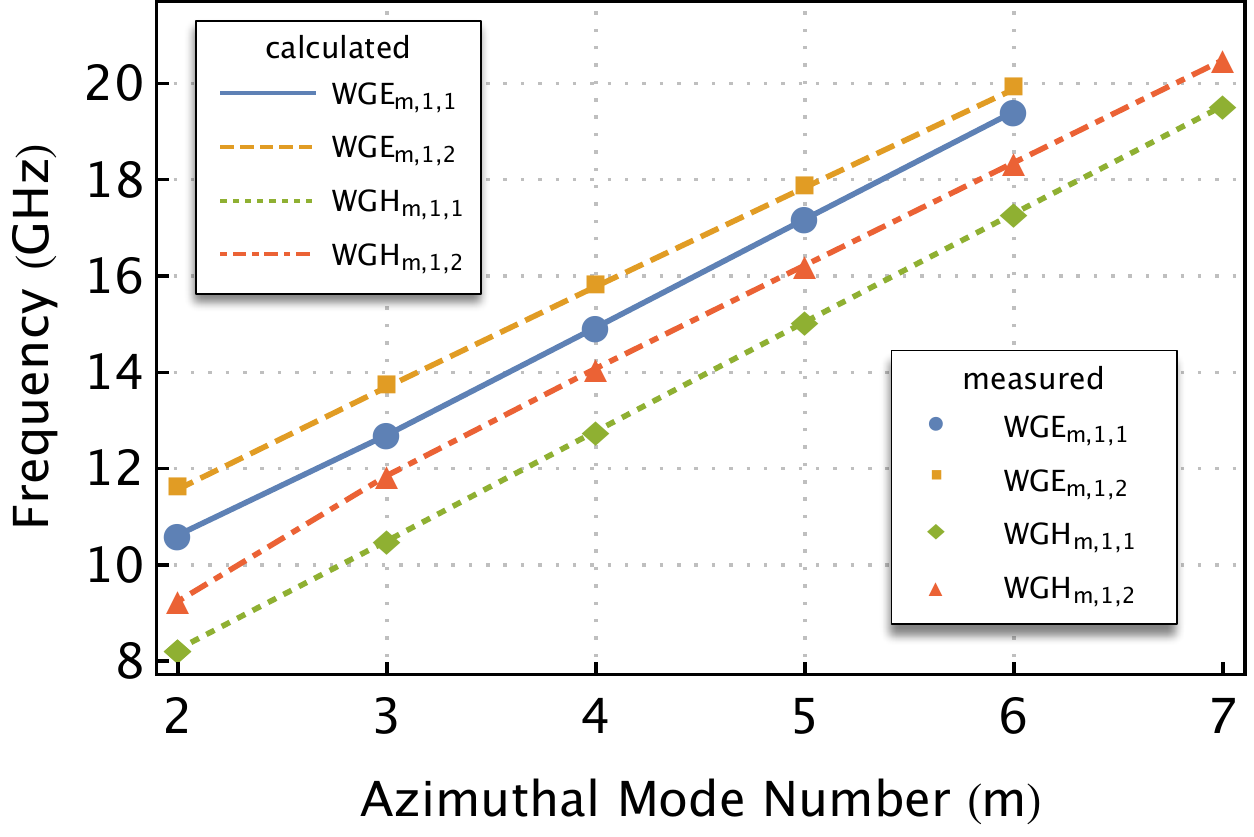}
\caption{Measured and calculated mode frequencies of the fundamental WGH (quasi-transverse magnetic) and WGE (quasi-transverse electric) mode frequencies with $\epsilon_r=14.488$.}
\label{fvsm}
\end{figure}

To further refine the model, a large number of resonant modes were measured with a vector network analyzer, and identified by comparison with the frequencies calculated in COMSOL. The highest $Q$-factor modes were exhibited by the fundamental WG mode families, which allowed easy identification. The frequency and $Q$-factors of the candidate modes will be discussed further below. The frequency discrepancy between measured and calculated values allowed us to determine more precisely the permittivity of our sample. For WG modes, most of the field is confined within the dielectric crystal, which greatly suppresses perturbations due to the metal support post and cavity walls. Correspondingly, the whispering gallery mode technique has been shown previously to be the most accurate for characterizing low-loss crystals\cite{jerzy2}. By comparing the four highest frequency WGE$_{m,1,1}$ and WGH$_{m,1,1}$ modes (as shown in Table \ref{WGmodes}) we determine the permittivity to be $\epsilon_r = 14.4879\pm0.0008$. This is the mean value of permittivity which results in a perfect match between simulation and experiment for the four modes, with the error defined as twice the standard error. This shows excellent relative precision, however, our precision in the dimensions of the cavity are actually much worse, and taking this into account results in a value of $\epsilon_r = 14.488\pm0.024$. Using this value, experiment and simulation are compared in Table \ref{WGmodes} and displayed graphically in Fig. \ref{fvsm}.


\begin{table}[ht]
\caption{Calculated and measured mode frequencies of the four modes used to calculate the permittivity}
\begin{center}
\begin{tabular}{|c|c|c|}
\hline
Mode & Calculated $f$ (GHz) & Measured $f$ (GHz)\\
\hline
WGH$_{7,1,1}$ & $19.5353\pm0.0195$ & 19.5350\\
WGE$_{6,1,1}$ & $19.4163\pm0.0194$ & 19.4160\\
WGH$_{6,1,1}$ & $17.3010\pm0.0173$ & 17.3010\\
WGE$_{5,1,1}$ & $17.1831\pm0.0172$ & 17.1840\\
\hline
\end{tabular}
\end{center}
\label{WGmodes}
\end{table}%

\begin{figure}[b]
\includegraphics[width=0.9\columnwidth]{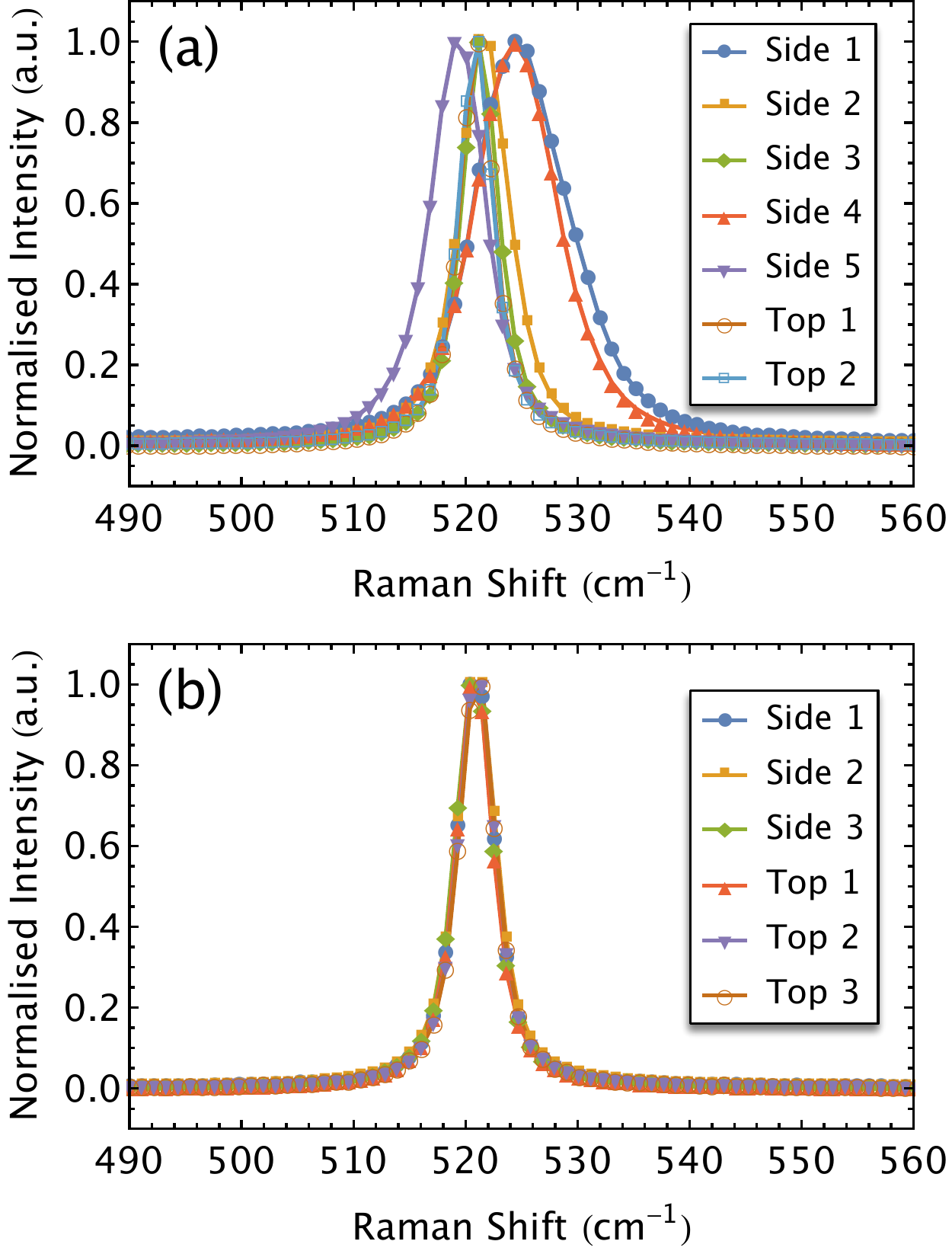}
\caption{Raman spectra showing the first order peak of silicon at 520 cm$^{-1}$ for one of the WG mode resonators pictured in Fig. \ref{isopure}. (a) Spectra measured at several locations on the side and top surfaces of the resonator after machining, and (b) after undergoing the acid treatment and annealing procedure. Surface strain is evident in the linewidth broadening and shifting of the Raman peak.}
\label{Raman}
\end{figure}

\section{Characterization and repair of surface strain}
The process of machining the resonators into uniform cylinders resulted in a significant degradation of the microwave $Q$-factors achievable in the sample. To investigate the effect of damage introduced by the lathe turning process, the surfaces of the samples were analysed using Raman spectroscopy with results shown in Fig. \ref{Raman}.  Raman spectra were acquired by focusing the 532 nm line of a frequency doubled Nd:YAG laser onto the surface of the resonator with a confocal microscope. A 100$\times$ objective lens with a numerical aperture of 0.9 resulting in a \hbox{$\sim$1 $\mu$m} laser spot size was used, with the laser power set to 15~mW (as measured before the objective) for all measurements. Spectra were collected for 10 seconds with a Renishaw InVia Reflex 0.25 working distance micro-Raman spectrometer with a 2400 grooves/mm grating. The overall wave number error was estimated to be \hbox{$\pm$0.5 cm$^{-1}$.} The spectral resolution of the setup was 1.1 cm$^{-1}$ per CCD pixel for the 2400 mm$^{-1}$ grating. 

Crystalline silicon has a characteristic Raman peak at 520 cm$^{-1}$ corresponding to inelastic scattering from the triply degenerate first-order optical phonon modes of silicon (one longitudinal and two transverse optic phonons). Strain or crystal imperfections can introduce linewidth broadening or shifting of this Raman peak. Upon examination, the end faces which were optically polished showed no such signs of surface strain, with the spectra in several locations nominally identical to that of a calibration sample of (natural abundance) silicon. However, the cylindrical side faces of the resonator exhibited strong variation in the Raman spectrum at several locations, indicating localised strain damage (Fig. \ref{Raman}(a)). This type of surface damage is known to introduce microwave losses \cite{surfaceloss,braginsky}, and is a likely cause of the low $Q$-factor values shown in grey in Fig. \ref{Qfafter}.

To repair the surface damage introduced by machining, a cleaning, acid etching, and annealing procedure was carried out. The $^{28}$Si resonator was first cleaned by sonication in acetone, isopropyl alcohol and de-ionised water for 10 minutes each. This was followed by 10 minute clean in Piranha solution (H$_2$SO$_4$(70\%) and H$_2$O$_2$(30\%) in a 4:1 ratio) and RCA solution (H$_2$O, H$_2$O$_2$(30\%), and HCl(30\%) in a 5:1:1 ratio) for 10 minutes. The resonator was then submerged in a 5\% solution of hydrofluoric acid in deionised water for 20 seconds, and rinsed with deionised water.

An anneal was then performed to relieve surface strain and damage, but additionally to reduce the density of paramagnetic centres on the silicon surface which may lead to increased microwave losses. This was achieved by annealing in oxygen for 1 hour at 1000$^{\circ}$C, followed by a 1 hour anneal in nitrogen to ensure a low fixed oxide charge density in the surface oxide. A forming gas anneal (5\% H in Ar) at 450$^{\circ}$C for 30 minutes was then used to passivate dangling bonds at the Si/SiO$_2$ interface. The density of states at the surface is estimated to be in the low 10$^{10}$ cm$^{-2}$/eV for the Si$\langle 100 \rangle$-SiO$_2$ interface as determined with electrical measurements of a test sample.

After undergoing these treatments, Raman spectra taken at several locations on the crystal indicated repair of surface strain, with no broadening or shifting of the characteristic Raman peak observed  (Fig. \ref{Raman}(b)).  The background corrected Raman data are well modelled using a Voigt function. Fits to the spectra from the as-machined surface had a mean Raman shift of $522.35$ cm$^{-1}$ (standard deviation $\sigma = 2.37$ cm$^{-1}$), with a mean FWHM linewidth of $11.71$ cm$^{-1}$ ($\sigma =6.21$ cm$^{-1}$), compared to a mean Raman shift of $520.95$ cm$^{-1}$ ($\sigma = 0.06$ cm$^{-1}$) and FWHM $2.93$ cm$^{-1}$ ($\sigma =0.06$ cm$^{-1}$) for the top surfaces. Such a shift is equivalent to a uniaxial strain of over 0.5~GPa.\cite{deWolf} After the acid treatment and annealing, the Raman data shows good recovery of the crystal, with mean Raman shift $520.9$ cm$^{-1}$ ($\sigma = 0.1$ cm$^{-1}$), and FWHM $3.92$ cm$^{-1}$ ($\sigma =0.14$ cm$^{-1}$).   

\section{Measurement after surface repair}

\begin{figure*}[t]
\includegraphics[width=0.7\textwidth]{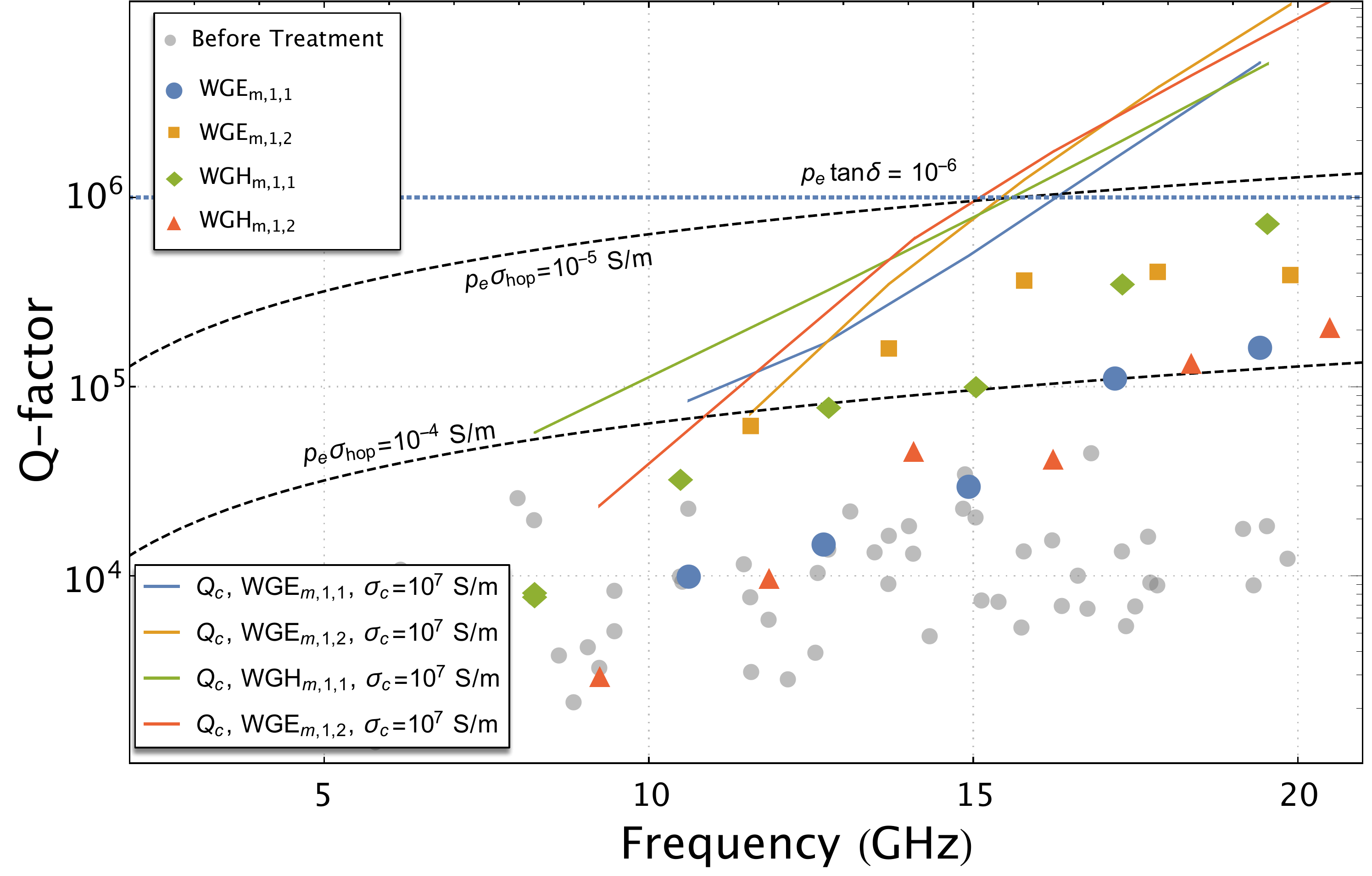}
\caption{Measured $Q$-factors of the resonator before (grey) and after (coloured) treatment, ensuring enough incident power to obtain maximum $Q$-factor. Additionally, potential $Q$-factors assuming dominant limiting loss factors are plotted for cavity losses; $Q_c$, assuming a cavity wall conductivity of $\sigma_c=10^{7}$ S/m, hopping conductivity; plotted for $p_e\sigma_\text{hop}=10^{-4}$ S/m and $p_e\sigma_\text{hop}=10^{-5}$ S/m, as well as pure dielectric loss assuming a loss tangent filling factor product $p_e \tan\delta=10^{-6}$.}
\label{Qfafter}
\end{figure*}

After processing to relieve surface strain and passivate dangling bonds, the $^{28}$Si resonator was placed in an aluminium cavity for cryogenic measurement. Aluminium becomes superconducting below 1.2 K, and thus was used in order to reduce microwave losses in the cavity walls when cooled to millikelvin temperature. In this way, the intrinsic $Q$-factor of the crystal can be determined with the highest accuracy, with minimal influence from other sources of loss such as resistive dissipation in the cavity walls. The measured $Q$-factors before (grey points) and after treatment (coloured points) are plotted in Fig. \ref{Qfafter}. 

The results show that the $Q$-factors after treatment have been restored close to the high values prior to machining \cite{nikita}. The results indicate that the maximum $Q$-factor was not fully reached given the data in Fig. \ref{Qfafter} appears as though $Q$-factors would continue increasing for higher frequency modes. 

Included in Fig. \ref{Qfafter} are the limits on $Q$-factors imposed by potential loss mechanisms. The total $Q$-factor of a resonance can be expressed as
\begin{equation}
	\frac{1}{Q}=\sum\frac{1}{Q_i},
	\end{equation}
where $1/Q_i$ are the losses due to the various mechanisms $i$, and therefore that which imposes the lowest $Q$-factor determines the final value. Losses in the metallic cavity walls can be estimated from the Geometric Factor of each mode, $G$, where $Q_c^{-1}=R_s/G$. Here $R_s$ is the surface resistivity of the cavity walls at a given frequency and
\begin{equation}
G=\omega\frac{\iiint_V\mu_0|\textbf{H}|^2dv}{\iint_S|\textbf{H}_\tau|ds},
\end{equation}
where $S$ is the internal surface of the cavity, $V$ is the total cavity volume including the silicon sample, and $\textbf{H}_\tau$ is the component of magnetic field tangential to $S$. The above equation can be calculated for each mode using the COMSOL finite element model, and the limit this loss mechanism places on mode $Q$-factors are plotted in Fig. \ref{Qfafter}, assuming surface conductivity of $\sigma_c=10^7$ S/m. This is an order of magnitude assumption chosen to approximate the measured data, and agrees within an order of magnitude of previously measured values \cite{al_cond}. We observe that at high frequencies this loss mechanism ceases to be dominant, but it is likely the dominant mechanism at low frequencies. 

Another loss mechanism theorised in Fig. \ref{Qfafter} is the conductivity of the silicon itself, or rather the hopping conductivity, $\sigma_\text{hop}$ \cite{jerzy}. At low temperatures, once electrons have been frozen out of the conduction band in a semi-conductor, a small amount of conductivity can remain due to electrons associated with impurities. These impurities are in very low concentrations and are unevenly distributed throughout the crystal, but the tail ends of their wavefunctions can overlap, and allow an electron to jump -- or {``}hop{''} -- from one location to another, resulting in conductivity and hence microwave losses \cite{book}. The frequency dependence of the conduction losses is
\begin{equation}
\frac{1}{Q_\text{cond}}=p_e\frac{\sigma}{\omega\varepsilon_0\varepsilon_r},
\end{equation}
where $\sigma$ is the conductivity of the resonator, and $p_e$ the electrical filling factor; a measure of the proportion of total electric field contained within the resonator \cite{jerzy}. Two limites in $Q$-factor due to hopping conductivity are plotted in Fig. \ref{Qfafter} for values of $p_e \sigma_\text{hop}=10^{-4}$ and $10^{-5}$ S/m. For high frequency modes, $p_e\sim1$. We see that the former would not permit the $Q$-factors measured for \hbox{$f>12$ GHz}, and therefore the hopping conductivity must be $10^{-5}$ S/m or larger, which would correspond to impurity concentrations less than $4\times10^{13}$ cm$^{-3}$ \cite{cond}, which agrees with the quoted concentration from the manufacturer. We can conclude from the frequency dependence of the measured $Q$-factors that this loss mechanism may dominate at higher frequencies.

An alternative loss mechanism operative at higher frequencies is pure dielectric loss from electronic and ionic polarization. If this were the case, then the silicon sample studied would have a pure dielectric loss tangent of approximately $\tan\delta\approx10^{-6}$. To adequately determine the dominant loss mechanism in this frequency regime, more measurements would need to be taken at \hbox{$f>20$ GHz.} 

\begin{figure}[t]
\includegraphics[width=\columnwidth]{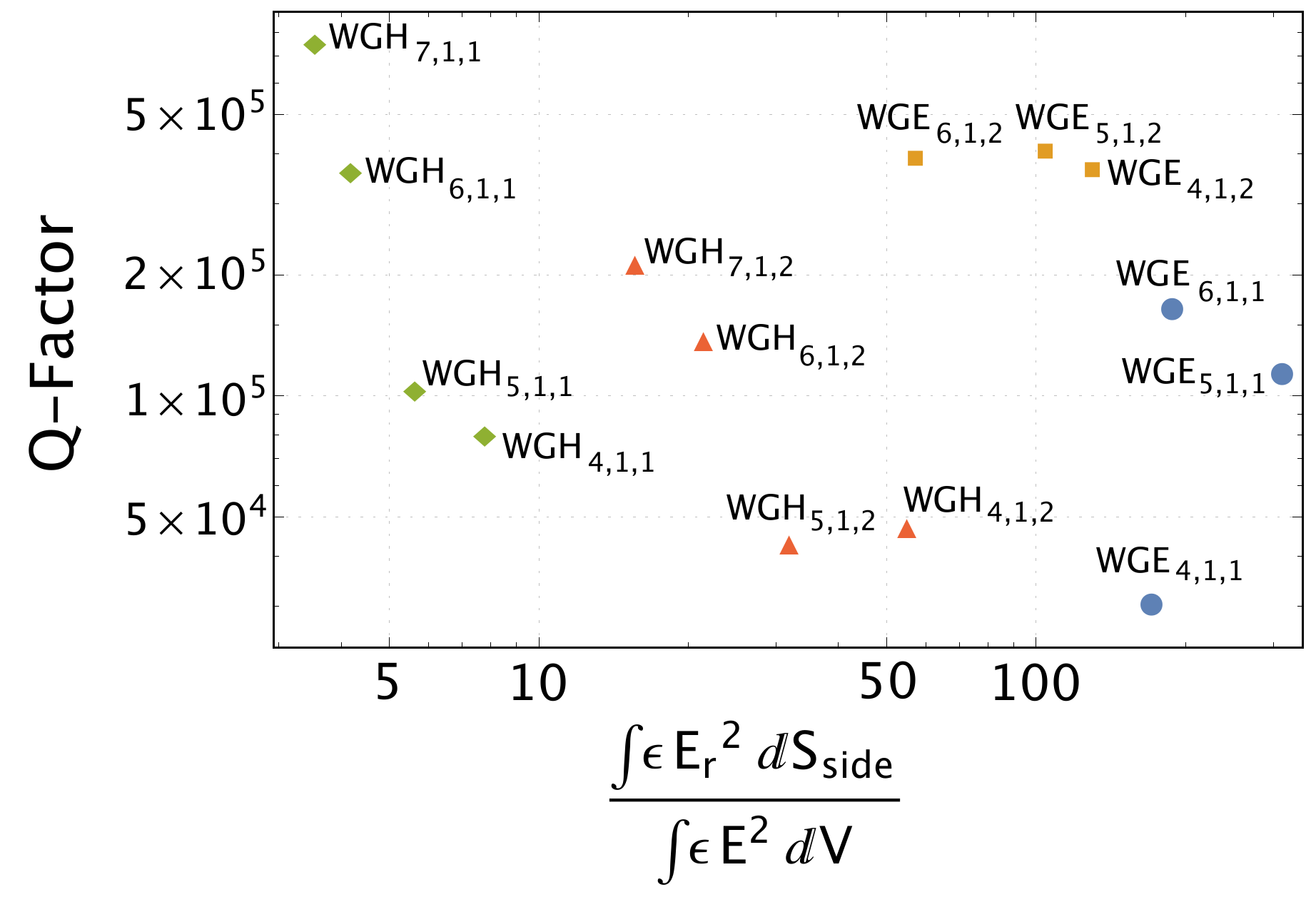}
\caption{Measured $Q$-factors categorised into mode families, plotted against surface integral of normal $E$ field of the curved surface of the cylinder. The apparent inverse relationship explains the consistent hierarchy of $Q$-factors for each mode family.}
\label{surface}
\end{figure}

One can not rule out that losses in this temperature range may also be attributed to nonuniform dopant distributions in the sample, or the accumulation of charge carriers on the surface of the sample \cite{jerzy}, mechanisms that are more difficult to model. In regards to the latter, a basic calculation of the proportion of electric field normal to the cylindrical surface of the sample at this boundary, formally calculated as $\iint_S\varepsilon|E_r|^2ds/\iiint_V \varepsilon|\textbf{E}|^2dV$, suggests that the observed hierarchy of mode $Q$-factors, specifically $Q_{\text{WGE}_{m,1,2}}>Q_{\text{WGE}_{m,1,1}}$ and $Q_{\text{WGH}_{m,1,1}}>Q_{\text{WGH}_{m,1,2}}$, may be attributed to this parameter. That is, the more electric field leaving the crystal surface radially, the larger the losses potentially due to surface charges. Fig. \ref{surface} shows this parameter versus $Q$-factor, and indeed we see that at least within the same polarisation of mode (E or H), there is a correlation between proportion of surface radial $\vec{E}$-field and $Q$. If losses were arising from some anisotropy in the crystal such as nonuniform conductivity or dopant distribution, one would not expect the orderly relationship observed between $Q$-factor and frequency.

\section{Conclusion}
The construction of WG mode resonators from isotopically pure silicon is a promising research direction given the enhanced $Q$-factors (and hence coherence time) of such electromagnetic resonances. We have successfully demonstrated the construction of these resonators and a technique to recover the high $Q$-factors post machining, which were initially lost due to surface strain imposed on the crystal and revealed through Raman spectroscopy. This result is crucial for the use of such resonators for hybrid quantum systems and frequency conversion applications, given silicon-28 has very low phonon losses, can host very narrow linewidth spin ensembles and is a material commonly used in optical applications. The next step will be to purposefully implant impurity ions and heal the implantation damage to recover the silicon lattice. In this was, narrow linewidth spin ensembles with clock transitions may be realised, which will couple to high-$Q$ WG modes inside the crystal.

\section{Acknowledgement}
This work was supported by Australian Research Council grant numbers CE170100009, CE170100012, a Research Collaboration Award from the University of Western Australia and the Defence Next Generation Technology Program.

\bibliography{Si_bib}

\end{document}